# Design of Rim-Located Reconfigurable Reflectarrays for Interference Mitigation in Reflector Antennas

Jordan Budhu, *Member, IEEE*, Sean V. Hum, *Senior Member, IEEE*, Steven Ellingson, *Senior Member, IEEE*, R. Michael Buehrer, *Fellow, IEEE*

*Abstract*— Radio telescopes are susceptible to interference arriving through its sidelobes. If a reflector antenna could be retrofitted with an adaptive null steering system, it could potentially mitigate this interference. The design of a reflectarray which can be used to reconfigure a radio telescope's radiation pattern by driving a null to the angle of incoming interference is presented. The reflectarray occupies only a portion of the rim of the original reflector and lays conformal to the paraboloid within this region. The conformal reflectarray contains unit cells with 1-bit reconfigurability stemming from two symmetrically placed PIN diodes. It is found that the dielectric and switch losses introduced by the reflectarray do not significantly affect the radio telescopes efficiency since the reflectarray is placed only along the outer rim of the reflector which is weakly illuminated. Simulation results of an L-band reconfigurable reflectarray for an 18m prime focus fed parabola are presented.

## I. Introduction

Radio telescopes are susceptible to interference arriving through its sidelobes [1]. If the reflector could adaptively steer nulls in the directions of the interfering signals, the interference could be mitigated. One technique to do this using an array of feeds placed in the focal plane has been proposed in [2]–[4]. Focal plane arrays have the drawback that mutual coupling changes with frequency [5] adding complexity to the calibration of the spatial filtering algorithms used for mitigation in broadband radio telescopes [6]. Another approach employs reconfigurable intelligent surfaces (RIS) to cancel the interference in the radio telescope aperture before reception [7]. The RIS is placed near the radio telescope and receives the same incident wavefront. The wavefront scattered by the RIS adds to the incident wavefront in the aperture of the radio telescope. The RIS wavefront is designed such that it destructively interferes with the interference in the incident wavefront in the aperture of the radio telescope thereby removing it. This approach however, literally redirects interference into the system, and relies on precise control over the phase and amplitude of the RIS scattered wavefront. Thus, the system is susceptible to element adjustment errors, and in the event of RIS failure, the system performance would even degrade relative to that of the radio telescope alone. A more robust approach is to null the interference in the aperture without introducing additional interference into the aperture. An approach which accomplished this was introduced in [8]. It involves a reflectarray antenna placed along the rim of a reflector antenna (see Fig. 1 and Fig. 2). In this approach, the reflectarray would dynamically reconfigure its scattered pattern such that it is 180° out of phase with the reflector pattern at the desired null angle. Although the concept was outlined in [8], imposed currents were

Jordan Budhu, Steven Ellingson, and R. Michael Buehrer are with the Bradley Department of Electrical and Computer Engineering, Virginia Tech, Blacksburg, VA 24060 USA. (e-mail: jbudhu@vt.edu, ellingso@vt.edu, rbuehrer@vt.edu). Sean Hum is with the Edward S. Rogers Sr. Department of Electrical and Computer Engineering, University of Toronto, Toronto, ON M5S3G4, Canada (sean.hum@utoronto.ca).

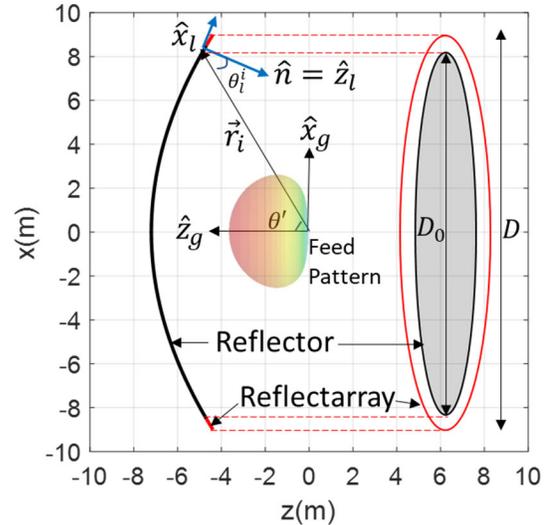

Fig. 1. Electronically reconfigurable reflectarray equipped high gain reflector antenna geometry.

used in the analysis. In this communication, a practical implementation is suggested. In particular, the detailed design and analysis of an electronically reconfigurable reflectarray antenna for interference mitigation in reflector antennas is presented realizing the concept in [8] using actual hardware.

Prior to this, in [9], [10], the concept was shown to work with flat non-reconfigurable reflectarray antennas. These configurations require additional support structure to affix the planar reflectarray to the paraboloidal reflector [10], or require an entirely new radio telescope design [9]. Instead, if a high gain reflector antenna could be retrofitted with a conformal reconfigurable reflectarray, then these systems would become adaptable with minimal changes to the original reflector geometry and positioning hardware. In this communication, an interference mitigation system (IMS) is designed that (a) can either be affixed to an existing radio telescope or designed into a new one, (b) does not require full-wave simulation or optimization in its design, and (c) can be reconfigured on-the-fly with appropriate weight selection algorithms.

As alluded to, the IMS consists of a reflectarray placed around the rim of the main reflector of an existing single or dual reflector radio telescope. In our case, the radio telescope is a prime focus fed paraboloidal reflector antenna (see Fig. 1). A portion around the rim of the original reflector of diameter $D$ is reallocated for reconfiguration. The remaining reflector now has a diameter of $D_0$. The rim area reallocated for reconfiguration ($D - D_0$) is chosen small enough such that the main beam maximum gain of the reflector is not significantly affected. Simultaneously, the reallocated rim area is chosen large enough to allow for sufficient scattered power from the reconfigurable reflectarray to place nulls in a desired direction. Fig. 3 illustrates this concept. The rim area is allocated large enough such that the pattern of the reflectarray, which contains its peak in the null direction, has a peak value equal to the pattern of the reflector at the desired null angle. Since the reflectarray pattern is designed to be 180° out of phase with the reflector pattern, the total pattern, reflector plus reflectarray, then attains a null in the desired direction. It was found in [8] that for a $D = 18$m reflector antenna, that $D_0 = 17$m met these requirements. The choice resulted in sufficient scattered power to place nulls in any direction within the angular range spanned by the sidelobes of the reflector while simultaneously not affecting the main lobe peak gain by more than a half of a dB. Subsequent work [11] has shown that even the main lobe gain change can be mitigated using constrained optimization techniques for selecting the element states. We will therefore start our designs at $D_0 = 17$m.

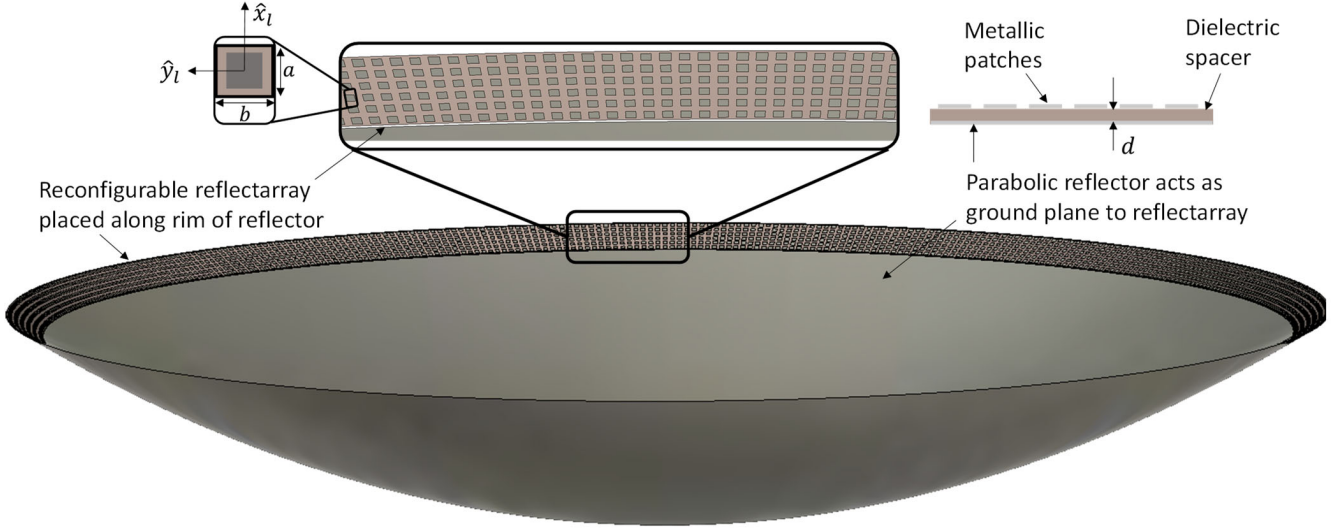

Fig. 2. IMS 3D model showing rim scattering reflectarray.

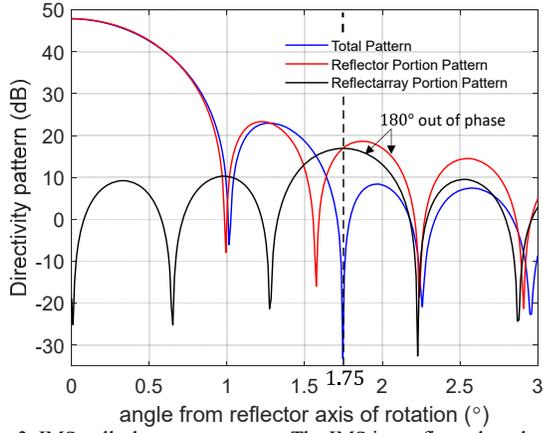

Fig. 3. IMS null-placement concept. The IMS is configured to place a null at the angle ($\theta_n^{-z} = 1.75°, \phi_n = 0°$).

Once the geometry is determined, the IMS is analyzed using integration of the physical optics (PO) current densities. The physical optics current densities are framed in terms of local reflection dyads. A reflection dyad can be viewed as a matrix containing reflection coefficients for all combinations of incident and reflected field polarizations. For the reflector portion, the reflection dyads are set equal to $-\bar{\bar{I}}$, where $\bar{\bar{I}}$ is the identity dyad. For the reflectarray portion, the reflection dyads are obtained using CST Microwave Studio (MWS) unit cell simulations for 1-bit reconfigurable unit cells (RUC). In this computational model, the unit cell is assumed to be part of an infinite planar array (IA) of like elements. The desired reflection dyads for null placement (switches 'on' or 'off') are found using a simple serial search method introduced in [8], although other more sophisticated approaches exist [11], [12]. Since both the PO approximation used over the reflector portion, and the IA approximation used over the reflectarray portion replace the local IMS geometry with an infinite tangent plane, several approximations must be first validated. These approximations include (a) the local periodicity (IA) approximation in the characterization of each unit cell, (b) tangent plane approximation for each unit cell printed on a curved surface, and (c) negligible effects of truncation at the edges of the parabolic reflectarray annulus. To validate the approximations, the reconfigurable reflectarray is replaced by a fixed (non-reconfigurable) reflectarray based on variable patch size and a full wave simulation is run. This simulation analyzes the actual geometry (see Fig. 2) with no approximations. The full-wave results are compared to that of the analysis approach utilizing the approximations presented in this communication. It is shown that the approximations used are reasonable. After validating the approximations, it is then shown that a reconfigurable reflectarray is capable of placing a null in a desired direction within the sidelobe envelope of the IMS pattern.

In Section II, the IMS geometry is defined. In Section III, the design and analysis algorithm of the IMS is presented. In Section IV, results are provided for a reflector antenna outfitted with a reconfigurable L-band reconfigurable reflectarray. An $e^{j\omega t}$ time convention is used and suppressed throughout.

## II. IMS GEOMETRY

The IMS is shown in Fig. 1. A reconfigurable reflectarray is placed along the rim of a prime focus paraboloidal reflector antenna with diameter $D$ and focal length $F$. In the projected aperture, the reflector portion has a diameter of $D_0$, and thus, the reflectarray occupies an annulus of width $(D - D_0)/2$. The reflector portion subtends an angle of $\theta' = \theta_1$ where $\theta_1 = 2\tan^{-1}(D_0/4F)$, and the reflectarray subtends the angles $\theta_1 \leq \theta' \leq \theta_0$ where $\theta_0 = 2\tan^{-1}(D/4F)$. The reflector is parameterized in the global coordinate system with axes denoted as $(\hat{x}_g, \hat{y}_g, \hat{z}_g)$ as

$$\vec{r}_i = F\sec^2\left(\frac{\theta'}{2}\right)\hat{r}, \quad \theta' \leq \theta_0 \tag{1}$$

A feed is placed at the global coordinate origin.

Physical optics is used to model the induced surface current density on the reflector antenna. In this approximation, at the point of reflection, the parabolic reflector is replaced by an infinite perfectly conducting tangent plane. The same approach is adopted to design the reflectarray, only now the tangent plane is replaced by an infinite doubly periodic reflectarray of identical patches. To maintain the rotational symmetry of the reconfigurable reflector system, the local coordinate axes $(\hat{x}_l, \hat{y}_l, \hat{z}_l)$ of this tangent plane are chosen such that the incident wave vector, $\vec{k}^i$, lies in the local $\hat{x}_l\hat{z}_l$-plane, where the $\hat{z}_l$-axis is equal to the parabolic reflector normal, $\hat{n}$ [13]

$$\hat{z}_l = \hat{n} = -\frac{2x_g}{4F}\hat{x}_g - \frac{2y_g}{4F}\hat{y}_g - \hat{z}_g$$

$$\hat{y}_l = \frac{\hat{n} \times \vec{k}^i}{|\hat{n} \times \vec{k}^i|}, \quad \hat{x}_l = \frac{\hat{y}_l \times \hat{n}}{|\hat{y}_l \times \hat{n}|} \tag{2}$$

In this configuration, the local incident angles to each patch in the reflectarray are $(\theta_l^i = \theta'/2, \phi_l^i = \pi)$. This speeds up the calculation of the reflection dyads of each patch since, in general, a unique reflection dyad must be computed for each unique pair of local plane wave incident angles $(\theta_l^i, \phi_l^i)$. Since $\theta_l^i$ varies by only 1.5° throughout the reflectarray region, the local incident angles to every patch in the reflectarray can be approximated as $(\theta_l^i = 31.25°, \phi_l^i = 180°)$. Thus, the CST MWS unit cell simulation based calculation of the reflection dyads for the RUC's need only be calculated for this illumination angle.

The reflectarray geometry is shown in Fig. 2. The reflectarray region is broken into rectangular unit cells of dimension $a$ by $b$. Each unit cell contains, in general, a rectangular patch of dimension $L$ in the $x_l$-direction and $W$ in the $y_l$-direction supported by a grounded dielectric substrate of thickness $d$ and permittivity $\epsilon = \epsilon_0\epsilon_r(1 - j\tan\delta)$. The patches are loaded with PIN diodes. The ground plane of the unit cell coincides with the reflector surface such that the annulus shaped reflectarray can be affixed to an existing reflector antenna along a portion of its rim. The reconfigurable reflectarray will be designed such that it can steer a null to any location within the first few sidelobes of the parent reflector antenna. Having defined the geometry, the design and analysis of the IMS can then be outlined.

## III. IMS Design and Analysis

### A. Calculation of the Incident Field

The $\hat{y}_g$-polarized incident field is calculated following from the raised cosine feed model

$$\vec{E}^i(r_i, \theta', \phi') = E_0 \frac{e^{-jkr_i}}{r_i} \cos^q\theta' \frac{(\hat{\theta}\sin\phi' + \hat{\phi}\cos\phi')}{\sqrt{1 - \sin^2\theta'\sin^2\phi'}} \tag{3}$$

where $E_0$ is a complex constant coefficient, $k = k_0 = 2\pi/\lambda_0$ is the wavenumber of free space, and $q$ is a parameter controlling the directivity of the feed radiation pattern.

### B. Calculation of Scattered Far Field

The scattered field is calculated following from the radiation integral

$$\vec{E}^s(r, \theta, \phi) = -j\omega\mu \frac{e^{-jkr}}{4\pi r} \int_{\theta'=0}^{\theta_0}\int_{\phi'=0}^{2\pi} \vec{J}_s(\vec{r}_i) e^{jkr_i[\sin\theta'\sin\theta\cos(\phi'-\phi)+\cos\theta'\cos\theta]}$$

$$\cdot r_i^2 \sin\theta'\sec\left(\frac{\theta'}{2}\right)d\theta'd\phi' \tag{4}$$

where $\omega$ is the radian frequency, $\mu$ is the permeability of free space, and $\vec{J}_s(\vec{r}_i)$ is the surface current density on the antenna. Ludwig's third definition is applied to determine the co-polarized component of the far field pattern

$$\vec{E}^s_{co,cr}(\theta, \phi) = \bar{\bar{T}} \cdot \vec{E}^s \tag{5}$$

where the Ludwig's dyad in the third definition, $\bar{\bar{T}}$, is defined as [14]

$$\bar{\bar{T}}(\phi) = \begin{bmatrix} 0 & 0 & 0 \\ 0 & -\cos(\phi) & -\sin(\phi) \\ 0 & -\sin(\phi) & \cos(\phi) \end{bmatrix} \tag{6}$$

for $\hat{x}$-polarized feeds and

$$\bar{\bar{T}}(\phi) = \begin{bmatrix} 0 & 0 & 0 \\ 0 & -\sin(\phi) & \cos(\phi) \\ 0 & -\cos(\phi) & -\sin(\phi) \end{bmatrix} \tag{7}$$

for $\hat{y}$-polarized feeds. Note the negative sign in the second column of (6) and (7) with respect to the definition in [14] due to the reversal in the direction of the $\hat{\theta}$ unit vector in the lower hemisphere ($x_g, y_g, z_g < 0$) where the radiation pattern is being calculated. The co-polarized and cross-polarized directivity patterns are then found from

$$D_{co}(\theta,\phi) = \frac{U_{co}}{U_{avg}}, \quad U_{co} = \frac{r^2}{2\eta_0}\left|\vec{E}^s_{co}\right|^2, \quad U_{avg} = \frac{P_{rad}}{4\pi}$$

$$D_{cr}(\theta,\phi) = \frac{U_{cr}}{U_{avg}}, \quad U_{cr} = \frac{r^2}{2\eta_0}\left|\vec{E}^s_{cr}\right|^2, \quad U_{avg} = \frac{P_{rad}}{4\pi} \tag{8}$$

where the radiated power, $P_{rad}$, is approximated as the power of the raised cosine feed intercepted by the antenna

$$P_{rad} = E_0^2 \frac{2\pi}{2\eta_0(2q+1)}\left(1 - \cos^{2q+1}\theta_0\right) \tag{9}$$

and $\eta_0$ is the wave impedance of free space. Note, calculating the radiated power using (9) allows one to determine the directivity without having to calculate the full 3D pattern. Thus, to calculate the radiation pattern, one only needs to determine the surface current density.

### C. Calculation of Surface Current Density

The surface current density induced on the IMS is calculated using the physical optics approximation

$$\vec{J}_s = 2\hat{n} \times \vec{H}^i = 2\hat{n} \times \vec{H}^r$$

$$= 2\hat{n} \times \frac{\hat{r} \times \vec{E}^r}{\eta_0} = -\frac{2}{\eta_0}\hat{n} \times \hat{z}_g \times \bar{\bar{R}}^{tot} \cdot \vec{E}^i \tag{10}$$

where $\hat{n}$ is given in (2), and $\vec{H}^i, \vec{H}^r$ and $\vec{E}^i, \vec{E}^r$ are the incident and reflected electric and magnetic fields, respectively. For the perfectly conducting reflector surface, $\bar{\bar{R}}^{tot} = -\bar{\bar{I}}$. For the reflectarray region, $\bar{\bar{R}}^{tot}$ is a function of the bias states of the reconfigurable elements. The computation of $\bar{\bar{R}}^{tot}$ for reconfigurable reflectarray elements is given in the next section.

### D. Calculation of the Reflection Dyad of the RUC

For the reconfiguration, we were inspired by the design in [15]. The $a = \lambda_0/2$ by $b = \lambda_0/2$ RUC is shown in Fig. 4. The unit cell consists of a grounded Taconic TLX-8 of thickness $d_1 = 13.5$mm and relative permittivity $\epsilon_r = 2.55(1 − j0.0017)$ bonded to a Rogers RT/Duroid 5880 substrate of thickness $d_2 = 4.167$mm. On the top face of the Taconic substrate, a square patch antenna of dimension $L = W = 51$mm in printed. At the center of two of the four radiating edges, a Skyworks SMP-1320-040LF pin diode is used to shunt the surface current on the patch to the ground layer through two metallic vias when the switch is on, and allow the patch to resonate when the switch is off. To maintain isotropy, both switches are either on or off simultaneously. The diode model parameters used in the simulations were taken from the data sheet and are provided in Table I. An additional via connects the center of the patch to the bias network placed on the back of the Rogers RT/Duroid layer through a hole in the ground plane. The bias network consists of a radial stub RF choke and a dc bias line.

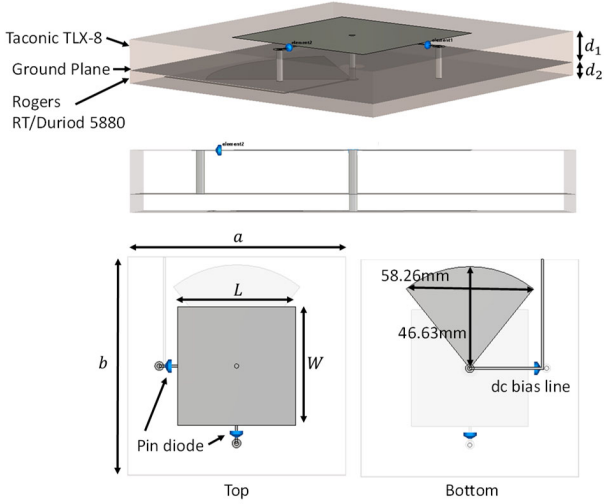

Fig. 4. Reconfigurable unit cell (RUC) design.

TABLE I: DIODE MODEL PARAMETERS

|  | $C_T$ (pF) | $R_s$ (Ω) | $L_s$ (nH) |
| --- | --- | --- | --- |
| On | 0 | 0.75 | 0.45 |
| Off | 0.23 | 0 | 0.45 |

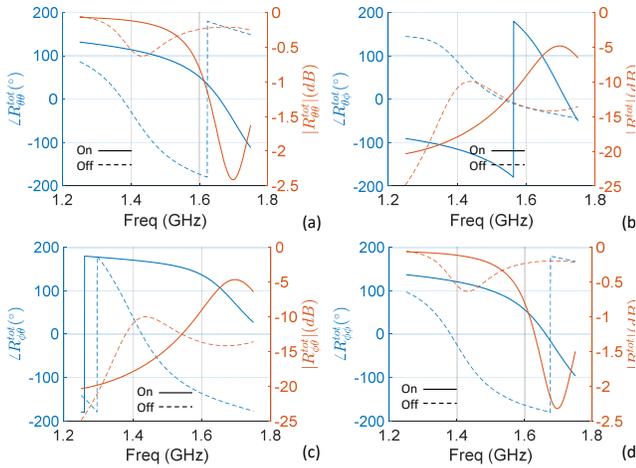

Fig. 5. Reflection dyad components for RUC for $\theta_l^i = 31.25°$, $\phi_l^i = 180°$. (a) $\angle R_{\theta\theta}(°)$ in blue and $|R_{\theta\theta}|$(dB) in red, (b) $\angle R_{\theta\phi}(°)$ in blue and $|R_{\theta\phi}|$(dB) in red, (c) $\angle R_{\phi\theta}(°)$ in blue and $|R_{\phi\theta}|$(dB) in red, and (d) $\angle R_{\phi\phi}(°)$ in blue and $|R_{\phi\phi}|$(dB) in red.

TABLE II: RUC REFLECTION DYAD

|  | DIODE ON | DIODE OFF |
| --- | --- | --- |
| $R_{\theta\theta}^{tot}$ | 0.97∠93.73° | 0.95∠ − 127.12° |
| $R_{\theta\phi}^{tot}$ | 0.185∠ − 146.18° | 0.274∠12.92° |
| $R_{\phi\theta}^{tot}$ | 0.187∠160.74° | 0.272∠ − 82.68° |
| $R_{\phi\phi}^{tot}$ | 0.97∠100.65° | 0.95∠ − 123.54° |

The reflection dyad, $\bar{\bar{R}}^{tot}$, computed using CST MWS unit cell analysis for the case of ($\theta_l^i = 31.25°$, $\phi_l^i = 180°$) and $f$=1.5GHz is shown in Fig. 5. As Fig. 5 shows, the RUC provides a 1-bit reconfiguration between the two-phase states of approximately $[−j, j]$ with a maximum loss of only 0.44dB. The RUC reflection dyad is tabulated in Table II. To design the IMS using the RUC, one needs now to simply determine which state ('on' or 'off') to set each RUC to in order to drive a null to a desired angle.

*E. Determination of Desired Reflection Dyads for Null Placement*

The desired reflection dyad for each patch, $\bar{\bar{R}}_{des}^{tot}$, to place a null at $(\theta = \theta_n, \phi = \phi_n)$ is found using a simple serial search introduced in [8]. The scattered far field in the direction of the desired null is first calculated from the induced surface current density on the reflector portion of the IMS over source angles $0 \leq \theta' \leq \theta_1$ and $0 \leq \phi' \leq 2\pi$ using (4). This gives the red pattern in Fig. 3. The reflectarray pattern (black pattern in Fig. 3) required to place a null in the desired direction is then synthesized patch-by-patch. Starting with the first patch in the reflectarray region, the scattered far field in the desired null direction is calculated for all members of a set of possible reflection coefficient states. For example, for both the 'on' and 'off' states of a 1-bit RUC (see Fig. 5 or Table II). The switch states which minimize the total co-polarized field magnitude (reflectarray plus reflector, blue pattern in Fig. 3) in the null direction are kept and the resultant field added to the cumulative total field sum. Then the contribution to the scattered far field from the next patch is evaluated. This process is continued until the desired reflection dyad for all patches is found.

*F. Calculation of IMS Efficiency*

The aperture efficiency, $\eta_{ap}$, of the IMS is defined here as the product of the spillover efficiency, $\eta_s$, the amplitude taper efficiency, $\eta_t$, and the radiation efficiency, $e_r$

$$\eta_{ap} = e_r \eta_s \eta_t \quad (11)$$

Assuming the maximum directivity occurs when $\eta_s \eta_t = 0.82$, the product of spillover and taper efficiency can be *approximated* as $\eta_s \eta_t = 0.731$ due to the drop in maximum directivity (~0.5dB, see next section) when the outer rim is reallocated for reconfiguration purposes. The radiation efficiency can be found from

$$e_r = \frac{\sum_1^N \left|\bar{\bar{R}}^{tot} \cdot \vec{E}^i\right|^2}{\sum_1^N \left|\vec{E}^i\right|^2} \quad (12)$$

where $N$ is the total number of reflection dyads. The expression in (12) can be interpreted as the ratio of the total power just after reflection to that just before. Hence, the dielectric, conductor, and RUC switch losses are included.

The gain of the IMS can then be found as

$$G = 10 \log\left[\eta_{ap} \frac{4\pi A}{\lambda^2}\right] \quad (13)$$

where $A$ is the IMS physical aperture area.

IV. RESULTS

As a first step, the radiation pattern of the fixed $D = 18$m reflector with focal length $F = 0.4D$ operated in the L-band at $f = 1.5$GHz is calculated using (3)-(10) for reference. The reflector subtends an angle of $\theta' = \theta_0 = 64.0°$. The $q$-factor for the feed model in (3) is set to a value of 1.14. The result is plotted in Fig. 6 along with a CST MWS verification. The co-pol directivity for this case is 48.38 dB and represents a reference value to measure the effects of the IMS system on the original radio telescopes directivity. For this case, $e_r$ calculated using (12) give a result of unity, and hence $\eta_{ap} = 0.82$ indicating $G = D_{co}$ by (13).

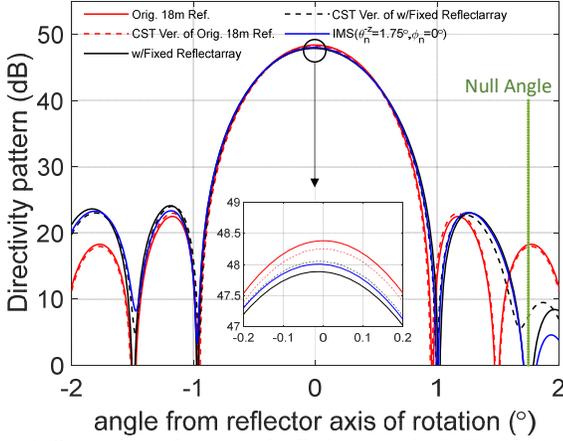

Fig. 6. Comparison of scattered far field patterns in the H-plane ($\phi = 0°$) for null angle ($\theta_n^{-z} = 1.75°, \phi_n = 0°$). The solid red curve represents the original $D_0$=18m reflector with no reflectarray and its CST MWS verification shown in the dashed curve. The solid black curve represents the $D_0$=17m reflector with non-reconfigurable reflectarray based on fixed square patches and its CST MWS verification shown in the dashed curve. The solid blue curve is the $D_0$=17m reflector with reconfigurable reflectarray (the IMS).

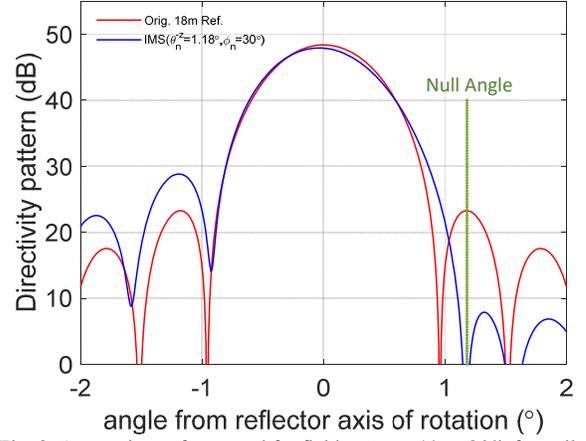

Fig. 8. Comparison of scattered far field patterns ($\phi = 30°$) for null angle ($\theta_n^{-z} = 1.18°, \phi_n = 30°$). The solid red curve represents the original $D_0$=18m reflector with no reflectarray. The solid blue curve is the $D_0$=16.5m reflector with reconfigurable reflectarray (the IMS).

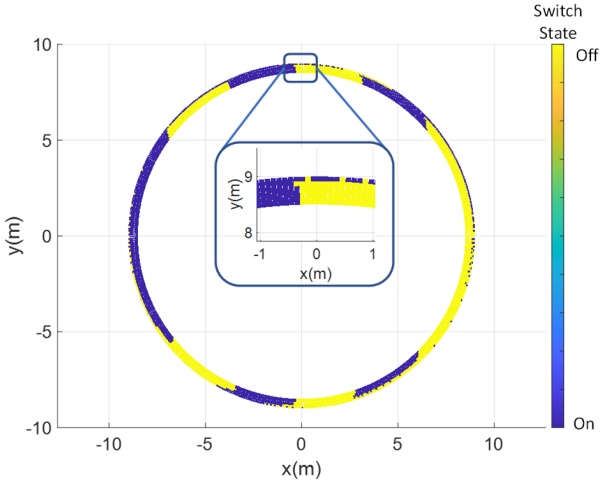

Fig. 7. Switch states of RUC's of IMS ($\theta_n^{-z} = 1.75°, \phi_n = 0°$).

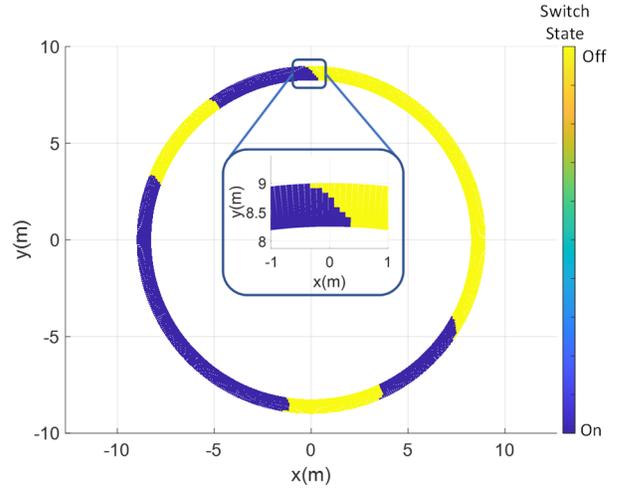

Fig. 9. Switch states of RUC's of IMS ($\theta_n^{-z} = 1.18°, \phi_n = 30°$).

To validate the IA tangent plane approximations used in the proposed algorithm, a non-reconfigurable (fixed) reflectarray based on variable patch size was designed to drive a null to the angle ($\theta_n = 181.75°, \phi_n = 0°$). Note, the null angle with respect to the reflector axis of rotation (the $-\hat{z}_g$-axis) is ($\theta_n^{-z} = 1.75°, \phi_n = 0°$). $D_0$ is chosen as 17m. In this design, the reflection dyads for the reflectarray region are computed using the Spectral Domain Method of Moments (SDMoM) [16]–[19] for a range of square patch dimensions. The same substrate and unit cell dimensions as that in Section III.D is used. First the serial search method is used to find the required reflection dyad for each unit cell from the set $\bar{\bar{R}}_{des}^{tot} = \pm j\bar{\bar{I}}$. Then the necessary patch dimensions are found by matching up the co-pol phase of the reflection dyads obtained using SDMoM to those found using the serial search method. In Fig. 6, the far field pattern resulting from the design algorithm of this communication which employs the approximations is compared to the full-wave CST MWS result which does not. Good agreement is obtained validating the IA tangent plane approximation used for the conformal reflectarray design of this communication. The directivity of this pattern is 47.88 dB representing a reduction of 0.5dB due to the reflectarray. Note, $e_r = 0.99$ for this case since the lossy materials are placed only along the outer rim which is weakly illuminated by the feed.

Having validated the IA tangent plane approximation, the unit cell simulations of the RUC's in Fig. 5 and Table II can be used in the algorithm to analyze the null driving performance of the IMS. The serial search method was applied to determine the switch states necessary to drive a null to the same angle ($\theta_n = 181.75°, \phi_n = 0°$). The resulting switch states are shown in Fig. 7. The switch states follow similar structuring as the imposed PO currents in [8] suggesting a possible fast reconfiguration approach through clustering of switches into contiguous groups. Note, since the reflectarray was sized to ensure enough scattered power to cancel the first sidelobe level [8], not all reflectarray elements are needed here to place a null in the second sidelobe. Hence the outermost ring or RUC's configure themselves to cancel their own radiation. Finally, the IMS pattern is calculated using the data in Fig. 5 and Table II in (10). The pattern is shown in Fig. 6. The directivity of the IMS pattern is 48.04 dB representing only a 0.34dB difference with respect to the original radio telescope pattern. Note, in this case, the pattern of the reconfigurable reflectarray in the direction of the main beam peak of the reflector adds in phase to the main beam of the reflector whereas in the fixed non-reconfigurable reflectarray case it does not (see Fig. 3 along 0° direction for fixed non-reconfigurable reflectarray case). Hence, the directivity is higher with the IMS versus with the non-reconfigurable fixed reflectarray. With the serial search method, it is not possible to also constrain the reflectarray patterns in the main beam direction, however, using more sophisticated switch state selection algorithms [11], it can. Despite the losses present in Fig. 5 and Table II, the radiation efficiency is still near unity. This is again because most of the power in the feed is not directed toward the lossy materials. Thus, the IMS with 1-bit RUC's can effectively place a null in a desired location within the sidelobe envelope without significantly affecting the radiation efficiency of the radio telescope.

The IMS was reconfigured to place a null at the peak of the first sidelobe at angle $(\theta_n^{-z} = 1.18°, \phi_n = 30°)$ as shown in Fig. 8. Although [8] concluded that $D_0 = 17$m was sufficient to place a null in the first sidelobe peak, to ensure a deep null in this case, the reflectarray size was increased by decreasing $D_0$ to $D_0 = 16.5$m. It should be noted that [8] considered ideal currents whereas the present paper considers realistic unit cells with non-zero switch losses. All other parameters are the same as the previous example. Figure 8 shows the IMS has the capability to null even the first sidelobe which is 5.73dB higher than the second sidelobe nulled in the previous example. The directivity of the IMS pattern is 47.92 dB, only 0.12 dB less than the previous design which utilized a 67.63% smaller reflectarray and still only 0.42 dB less than the directivity of the original 18m radio telescope. Despite the larger reflectarray now containing 9 concentric paraboloidal rings of elements rather than the 6 of the previous design, the radiation efficiency is still near unity. The switch states are shown in Fig. 9. Comparing Fig. 9 to Fig. 7, the switch state pattern has rotated by 30° on account of the new azimuthal null angle. These two examples show the IMS has the capability to null any sidelobe over all azimuth angles and hence is a viable candidate for interference mitigation in radio telescopes.

## V. Conclusion

We have presented a practical design for the interference mitigation system (IMS) conceived in [8]. The IMS consists of a reconfigurable reflectarray antenna placed along the outer rim of a radio telescope's high gain main reflector antenna. The methodology for the design and analysis of the IMS has been presented, and results have been shown to be consistent with the commercial software CST MWS but with the following advantage due to the tangent plane approximation introduced in this communication: the analysis of the IMS requires 1TB less computer memory than the CST MWS simulation and runs over 100 times faster (for reference, the CST MWS simulation took 6 days, 18 hours, and 48 mins, and consumed 1.17TB of random access memory). The null forming results are consistent with [8] and [9], [10] thereby further confirming the efficacy of the concept. Also, because the lossy materials were placed only along the rim of the reflector where very little power is being directed from the feed, it was found that the substrate and pin diode switch losses did not affect the overall efficiency of the IMS. One-bit dual-band RUC's with wide bandwidth within each band are planned for future extensions to this work.

ACKNOWLEDGEMENT

This work was supported in part by National Science Foundation grant AST-2128506.

References

[1] International Telecommunication Union: Radio Astronomy Series, "ITU-R RA.2126-1: Techniques for mitigation of radio frequency interference in radio astronomy, https://www.itu.int/dms_pub/itu-r/opb/rep/R-REP-RA.2126-1-2013-PDF-E.pdf," 2013.
[2] J. Duggan and P. McLane, "Adaptive beamforming with a multiple beam antenna," in *ICC '98. 1998 IEEE International Conference on Communications. Conference Record. Affiliated with SUPERCOMM'98 (Cat. No.98CH36220)*, vol. 1, pp. 395–401.
[3] A. Pal, A. Mehta, A. Skippins, P. Spicer, and D. Mirshekar-Syahkal, "Novel Interference Suppression Null Steering Antenna System for High Precision Positioning," *IEEE Access*, vol. 8, pp. 77779–77787, 2020.
[4] C. K. Hansen, K. F. Warnick, B. D. Jeffs, J. R. Fisher, and R. Bradley, "Interference mitigation using a focal plane array," *Radio Sci*, vol. 40, no. 5, 2005.
[5] D. T. Emerson and J. M. Payne, Eds., *Multi-feed Systems for Radio Telescopes (Astronomical Society of the Pacific conference series)*. Astronomical Society of the pacific, 1995.
[6] K. F. Warnick and M. A. Jensen, "Effects of mutual coupling on interference mitigation with a focal plane array," *IEEE Trans Antennas Propag*, vol. 53, no. 8, pp. 2490–2498, Aug. 2005.
[7] Z. Zou, X. Wei, D. Saha, A. Dutta, and G. Hellbourg, "SCISRS: Signal Cancellation using Intelligent Surfaces for Radio Astronomy Services," in *2022 IEEE Global Communications Conference, GLOBECOM 2022 - Proceedings*, 2022.
[8] S. Ellingson and R. Sengupta, "Sidelobe Modification for Reflector Antennas by Electronically Reconfigurable Rim Scattering," *IEEE Antennas Wirel Propag Lett*, vol. 20, no. 6, pp. 1083–1087, Jun. 2021.
[9] S. V. Hum, S. W. Ellingson, and R. M. Buehrer, "Reflectarray Concept for Interference Mitigation in Radio Astronomy," in *IEEE Int'l Ant. & Prop. Sym.*, Portland OR, 2023.
[10] S. V. Hum, S. W. Ellingson, and R. M. Buehrer, "Toward Electronically Reconfigurable Rims for Reflectors in Radio Astronomy," in *URSI General Assembly & Sci. Sym.*, Sapporo, Japan, 2023.
[11] R. M. Buehrer and S. W. Ellingson, "Weight Selection for Pattern Control of Paraboloidal Reflector Antennas with Reconfigurable Rim Scattering," in *2023 IEEE Aerospace Conference*, 2023, pp. 1–8.
[12] R. M. Buehrer and S. W. Ellingson, "Pattern Control for Reflector Antennas Using Electronically-Reconfigurable Rim Scattering," in *2022 IEEE International Symposium on Antennas and Propagation and USNC-URSI Radio Science Meeting (AP-S/URSI)*, 2022, pp. 577–578.
[13] C. A. Balanis, *Antenna Theory: Analysis and Design*, 3rd ed. Hoboken, NJ: John Wiley & Sons, 2005.
[14] A. Ludwig, "The definition of cross polarization," *IEEE Trans Antennas Propag*, vol. 21, no. 1, pp. 116–119, Jan. 1973.
[15] H. Yang, F. Yang, S. Xu, M. Li, X. Cao, and J. Gao, "A 1-Bit Multipolarization Reflectarray Element for Reconfigurable Large-Aperture Antennas," *IEEE Antennas Wirel Propag Lett*, vol. 16, 2017.
[16] T. A. Metzler and D. H. Schaubert, "Design and analysis of a microstrip reflectarray," DIssertation, University of Massachusetts, Amherst, Massachusetts, 1993.
[17] D. M. Pozar, S. D. Targonski, and H. D. Syrigos, "Design of millimeter wave microstrip reflectarrays," *IEEE Trans Antennas Propag*, vol. 45, no. 2, pp. 287–296, Feb. 1997.
[18] S. R. Rengarajan, "Choice of basis functions for accurate characterization of infinite array of microstrip reflectarray elements," *IEEE Antennas Wirel Propag Lett*, vol. 4, pp. 47–50, 2005.
[19] R. Mittra, C. H. Chan, and T. Cwik, "Techniques for analyzing frequency selective surfaces-a review," *Proceedings of the IEEE*, vol. 76, no. 12, pp. 1593–1615, Dec. 1988.